\documentclass[aps,tightenlines,twocolumn,groupaddress]{revtex4-1}
\usepackage{graphicx}
\usepackage{dcolumn}
\usepackage{bm}
\usepackage{csquotes}
\usepackage{epigraph}
\usepackage{amssymb}
\usepackage{color,graphicx}
\usepackage{subfigure}
\usepackage{lmodern}
\usepackage{amsmath}
\usepackage{amsbsy}
\usepackage{amsthm}
\usepackage{float}
\usepackage{bbm}

\begin{document}

\title{Clausius inequality versus quantum coherence}

\author{Ali Soltanmanesh}
\email[]{soltanmanesh@ch.sharif.edu}
\author{Afshin Shafiee}
\email[Corresponding Author:~]{shafiee@sharif.edu}
\affiliation{Research Group on Foundations of Quantum Theory and Information,
Department of Chemistry, Sharif University of Technology
P.O.Box 11365-9516, Tehran, Iran}
\affiliation{School of Physics, Institute for Research in Fundamental Sciences (IPM), P.O.Box 19395-5531, Tehran, Iran}

\begin{abstract}
In this study, we model a harmonic oscillator that enters an interferometer partially coupled to a thermal bath of oscillatory fields by employing a Brownian-type Lindblad master equation. More specifically, we investigate the dynamics and the variations of the thermodynamic quantities of the system at different temperatures. We recognize that although the system can remain coherent during its interaction with the thermal bath in the low-temperature limit, the system's entropy production violates the Clausius inequality. Furthermore, we argue that the system's coherence is the source of this violation, rather than the entanglement degree of system-environment, as reported in previous studies.
\end{abstract}

\maketitle

\section{Introduction}

Thermodynamics, as a fundamental framework to investigate many fields of sciences, can always be employed to characterize physical processes. Since the inception of the scientific revolution of physics, thermodynamics has always been employed and now is applicable in a large variety of macroscopic systems  \cite{Gold,Liu}. Thermodynamics laws are well-known for closed and open classical systems. Nevertheless, the meaning of these laws and how they apply to the microscopic regime are still blurred  \cite{Gem,Goo1,Vin}. Several inspiring works have been conducted in this regard; e.g., defining the entropy and the second law of thermodynamics in the quantum domain \cite{Bra,Bin,Deu,Cwi}, investigating the fundamental role of information theory in quantum thermodynamics \cite{Goo1,Pop,del} and seeking for the meaning of work, heat, and energy transfer in micro-systems \cite{And,Kie,All,Esp}.

In deriving the Gibbs distributions for closed systems, the fundamental statistical assumptions are independent of the quantum or classical nature of the system \cite{Land}. Moreover, to study the open subsystems, we consider an interaction between the subsystem and a thermal bath. Under general statistical conditions, the thermal baths are quite identical regardless of classical or quantum approaches \cite{All2,Kli}. However, understanding the connection between the thermodynamics of these regions remained a controversial subject despite years of research. Crucial questions concerning the origins of probabilities and entropy in the microscopic regime are not satisfactorily resolved yet such that in different periods they have been interpreted proportional to the achievements of the date \cite{Pop,Eva,Sag,Jac,Groo,Gog}.

There is a general belief that open quantum systems are consistent with the laws of equilibrium thermodynamics \cite{Nie}. In this regard, decoherence theory is quite helpful to define the thermodynamic laws of open quantum systems. Moreover, one can study fluctuations of thermodynamic quantities of the system through its interaction with the environment \cite{Hen,Wei,Esp2,Esp3,Gem2,Wil,San}. For instance, Binder and co-workers formulated operational thermodynamics suitable for an open quantum system undergoing a general quantum process and presented it as a complete positive and trace-preserving (CPTP) map \cite{Bin}. Moreover, in another approach, thermodynamics was formulated based on the Clausius inequality with a proper definition of work and heat \cite{Nie}.  In these works, the thermodynamic description is based on a single open quantum system; however, there are alternative approaches in understanding thermodynamics in the microscopic regime. For example, the emergence of thermodynamic phenomena in \cite{Deu2,Rig} depends on the complexity of the spectrum of the large quantum systems. Furthermore, in recent studies, information has shown an essential role in describing thermodynamic events. In this connection, a vast variety of studies have been conducted based on quantum information  \cite{Goo1}; e.g., introducing entropy measures for quantifying uncertainties  \cite{Cov}, study of equilibration in quantum regime and maximum entropy principle  \cite{Lin,Sho}, the concept of thermalization in quantum thermodynamics  \cite{Rie}, the role of entanglement theory \cite{Hor1}, Landauer's erasure principle and information theory  \cite{Ple}, and the thermodynamics of information \cite{Par}.

Despite all the efforts made to determine thermodynamic quantities, especially entropy, in microscopic scales, there is no proper evidence to examine how these definitions work. Nonetheless, recently few but interesting studies shed light on the concept; e.g., the violation of the Clausius inequality for a quantum harmonic oscillator linearly coupled to a bath of oscillatory fields \cite{All2,Hil}, the comparison between the thermodynamic and von Neumann entropy of a quantum Brownian oscillator \cite{Hor}, and description of quantum coherence in thermodynamic processes \cite{Lost,Lost2}.

Furthermore, the study of quantum thermodynamic processes and understanding their intrinsic limitations is essential nowadays considering the developments in quantum thermal machines, quantum heat engines, and related technologies \cite{Kos,Gelb}. Quantum thermal machines perform useful tasks by applying thermal gradients to the environment, such as the production of work \cite{Quan,Roul}, the refrigeration of a quantum degree of freedom \cite{Levy,Mit}, the entanglement production \cite{Boy} and quantum clocks \cite{Erk}. In this regard, several valuable works have been done to investigate the boundaries of quantum thermodynamic processes in open systems \cite{Hof}.

In the present work, we assess the validity of the Clausius inequality for a harmonic oscillator passing through a Mach-Zehnder type interferometer that is partially coupled to a thermal bath. To our expectation, as the system evolves through time, it loses its coherent properties satisfying the Clausius inequality. However, at the low-temperature regime, the system's entropy begins to trespass the Clausius inequality. In addition, we examine how the system's coherence varies while the Clausius inequality is violated. Finally, we discuss the source of such an observation  \cite{All2,Hil}.

The remainder of this paper is organized as follows. In section II, we briefly review the Lindblad dynamics of a harmonic oscillator interacting with a thermal bath consisting of harmonic oscillators. In section III, we describe a harmonic oscillator, entering an interferometer which is partially coupled to a thermal bath. Also, we study the dynamics and the thermodynamic quantities of such a system. Finally, we inquire to keep the system's coherence during its interaction with the thermal bath at low temperatures, when the system?s entropy exceeds the Clausius inequality in section IV. Moreover, we discuss that the violation is not due to the amount of the system-environment entanglement, but to the system?s coherence. Finally, in section V, we conclude the paper.


\section{Lindblad Model of Quantum Brownian Motion}

The dynamics provided here are based on employing the model of the quantum Brownian motion introduced by Maniscalco et al., who applied the Gorini-Kossakowski-Sudarshan-Lindblad (GKSL) master equation \cite{Man,Gori}. In this system-bath model, the total Hamiltonian is defined as
\begin{equation}
\label{totalH}
\hat{H}=\hat{H}_s+\hat{H}_\varepsilon+\hat{H}_{int},
\end{equation}
where $\hat{H}_s$, $\hat{H}_\varepsilon$ and $\hat{H}_{int}$ are the Hamiltonians of the system, the bath (environment) and the system-bath interaction, respectively. We consider a harmonic oscillator as the central system and a thermal bath of the electromagnetic harmonic oscillators as the environment. Accordingly, the total Hamiltonian is written as 
\begin{equation}
\label{TH}
\hat{H}=\frac{1}{2}(\hat{P}^2+\Omega^2 \hat{X}^2)+\sum_i \frac{1}{2}(\hat{p}_i^2+\omega_i^2 \hat{q}_i^2)+\hat{H}_{int},
\end{equation}
where $\Omega$ ($\omega$) is the system (the environment) frequency, and $\hat{P}$ and $\hat{X}$ ($\hat{p}$ and $\hat{q}$) are momentum and position operators of the system (the environment), respectively. We removed the mass from the Eq. \eqref{TH} and considered $\hbar=1$. The interaction Hamiltonian reads as
\begin{equation}
\label{HInt}
\hat{H}_{int}=\hat{X}\otimes \hat{E}=\hat{X}\otimes\sum_i c_i \hat{q}_i,
\end{equation}
where the position coordinate of the central system  $\hat{X}$ linearly couples to the position $\hat{q}_i$ of the $i$-th thermal bath oscillator with the coupling strength $c_i$.  Here, $\hat{E}$ denotes the environment operator.

Considering $\hat{\rho}$ as the total density matrix, the following assumptions can be made in the order of their appearance. First, the system and the environment are supposed to be uncorrelated at  t=0; i.e.,  \mbox{$\hat{\rho}(0)=\hat{\rho}_s(0)\otimes\hat{\rho}_\varepsilon(0)$} where $\hat{\rho}_s$ and $\hat{\rho}_\varepsilon$ are the system and the environment density matrices, respectively. Second, we assume that the environment is stationary; i.e.,  $[\hat{H}_\varepsilon,\hat{\rho}_\varepsilon(0)]=0$, and also the expectation value of $\hat{E}$ is zero; i.e., $\text{Tr}_E [\hat{\rho}_E(0)] = 0$. Finally, the system-environment coupling is weak and we neglect the effect of the oscillator frequency renormalization since it is negligible under the weak coupling. Hence, by averaging over rapidly oscillating terms, one gets the following secular approximated master equation \cite{Man,Bre}
\begin{align}
\label{MaEq}
\frac{d\hat{\rho}_s}{dt}=&-\frac{\Delta(t)+\gamma(t)}{2}[\hat{a}^\dagger \hat{a}\hat{\rho}_s-2\hat{a}\hat{\rho}_s \hat{a}^\dagger+\hat{\rho}_s \hat{a}^\dagger \hat{a}] \nonumber \\
&-\frac{\Delta(t)-\gamma(t)}{2}[\hat{a}\hat{a}^\dagger \hat{\rho}_s-2\hat{a}^\dagger\hat{\rho}_s \hat{a}+\hat{\rho}_s \hat{a} \hat{a}^\dagger],
\end{align}
where $\hat{a}=(\hat{X}+i\hat{P})/\sqrt{2}$ and $\hat{a}^\dagger=(\hat{X}-i\hat{P})/\sqrt{2}$ are the bosonic annihilation and creation operators, respectively. Also, the time-dependent coefficients $\gamma(t)$ and $\Delta(t)$ represent the classical damping and diffusive terms, respectively. These coefficients are defined as
\begin{align}
\label{D}
\Delta(t)&=\int_0^t\kappa(\tau)\cos(\Omega\tau)\text{d}\tau, \\
\label{g}
\gamma(t)&=\int_0^t\mu(\tau)\sin(\Omega\tau)\text{d}\tau,
\end{align}
where
\begin{equation}
\label{noise}
\kappa(\tau)=\sum_ic_i^2\langle \{q_i(\tau),q_i\}\rangle,
\end{equation}
and
\begin{equation}
\label{dissipation}
\mu(\tau)=i\sum_ic_i^2\langle [q_i(\tau),q_i]\rangle,
\end{equation}
are noise and dissipation kernels, respectively.

The master equation \eqref{MaEq} is in the Lindblad form in situations, where the quantity $\Delta\pm\gamma$ remains positive in all times \cite{Man2}. Now we consider the case of an Ohmic spectral density for the bath with Lorentz-Drude cutoff \cite{Schl}
\begin{equation}
\label{sd}
J(\omega)=\frac{2\gamma_0\omega}{\pi}\frac{\Lambda^2}{\Lambda^2+\omega^2},
\end{equation}
where $\Lambda$ is the cut-off frequency and the dimensionless factor $\gamma_0$ describes the system-environment effective coupling strength. Thus, for the expressions of noise and dissipation kernels, we have
\begin{align}
\kappa(\tau)&=\sum_i\frac{c_i^2}{2\omega_i}\text{coth}(\frac{\omega_i}{2kT})\text{cos}(\omega_i\tau) \nonumber \\
&\equiv\int_0^\infty \text{d}\omega J(\omega)\text{coth}(\frac{\omega}{2kT})\text{cos}(\omega\tau) \\
\mu(\tau)&=\sum_i\frac{c_i^2}{2\omega_i}\text{sin}(\omega_i\tau)  \nonumber \\
&\equiv\int_0^\infty \text{d}\omega J(\omega)\text{sin}(\omega\tau),
\end{align}
where $k$ is the Boltzmann constant and $T$ denotes the temperature. At the asymptotic long-time limit, the coefficients $\Delta(t)$ and $\gamma(t)$ approach their stationary values. So, their expressions up to the second order of coupling constant read as \cite{Man}
\begin{align}
\label{Delta}
\Delta &=\gamma_0^2\Omega\frac{r^2}{1+r^2}\coth(\frac{\Omega}{2kT}) \\
\label{Gamma}
\gamma &=\gamma_0^2\Omega\frac{r^2}{1+r^2},
\end{align}
where $r=\Lambda/\Omega$. Therefore, the master equation (\ref{MaEq}) becomes similar to the well-known Markovian master equation of damped harmonic oscillator
\begin{align}
\label{ME}
\frac{d\hat{\rho}_s}{dt}=&-\Gamma(\bar{n}+1)[\hat{a}^\dagger \hat{a}\hat{\rho}_s-2\hat{a}\hat{\rho}_s \hat{a}^\dagger+\hat{\rho}_s \hat{a}^\dagger \hat{a}] \nonumber \\
&-\Gamma\bar{n}[\hat{a}\hat{a}^\dagger \hat{\rho}_s-2\hat{a}^\dagger\hat{\rho}_s \hat{a}+\hat{\rho}_s \hat{a} \hat{a}^\dagger],
\end{align}
where $\Gamma=\gamma_0^2\Omega r^2/(1+r^2)$ and $\bar{n}=(e^{\Omega/kT}-1)^{-1}$. The positiveness of the coefficients $\Delta\pm\gamma$ assures us that the master equation \eqref{MaEq} is in the Lindblad form.

\begin{figure}
\centering
\includegraphics[scale=0.36]{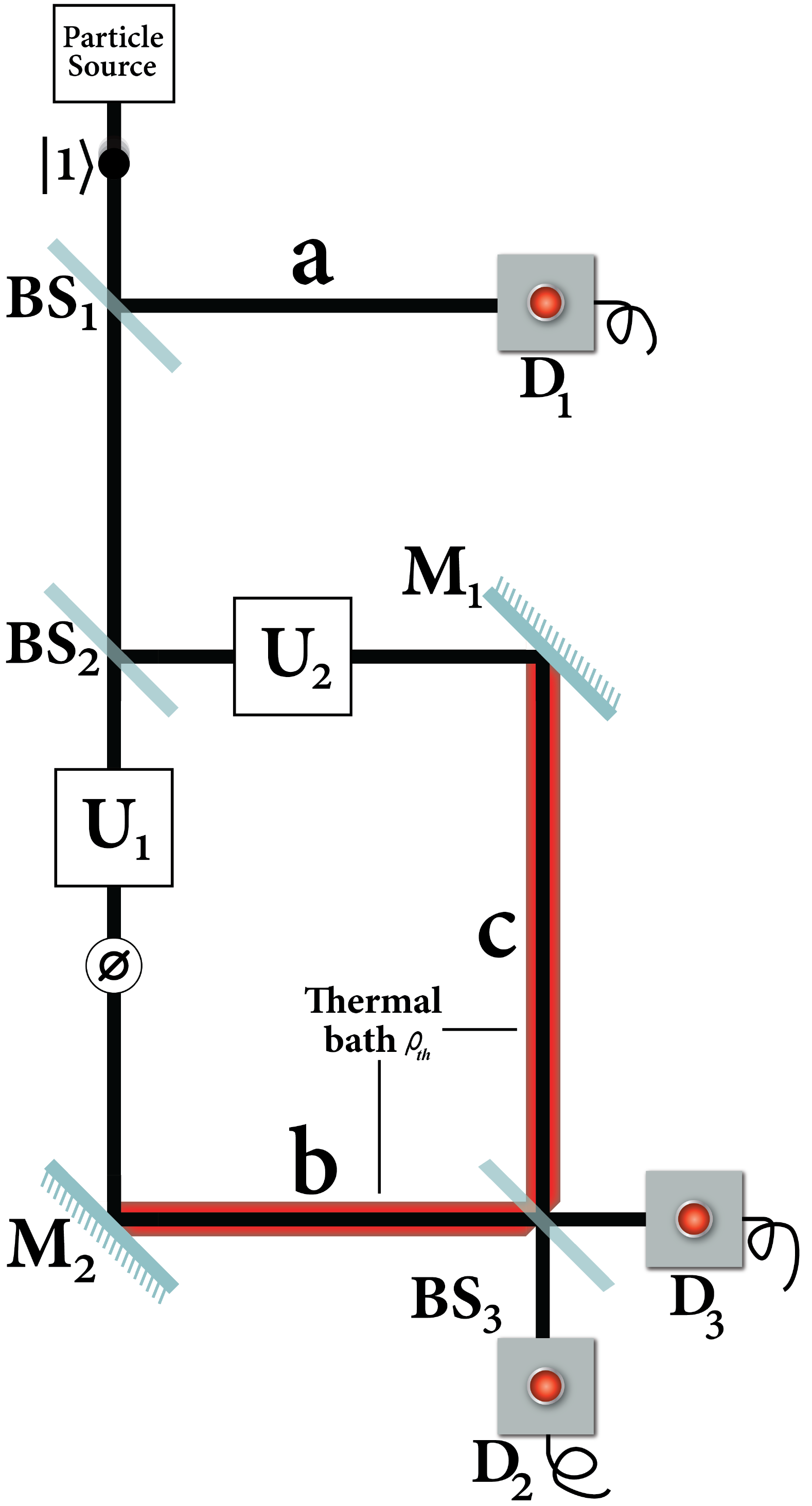}
\caption{A particle in the state $\vert 1\rangle$ enters the interferometer through the beamsplitter $BS_1$. Then, in the reflected path $(a)$, it reaches the detector $D_1$, and/or the particle passes through the $50$:$50$ beamsplitter $BS_2$. After reaching the unitary gates $U_1$ and $U_2$ in paths (b) and (c) respectively, and reflecting via the Mirrors $M_1$ or $M_2$, it interacts with the thermal bath, influenced by the decoherence process. Afterward, by passing through the beamsplitter $BS_3$, it reaches the detectors $D_2$ or $D_3$.}\label{inter}
\end{figure}

\section{An Interferometer Partially Coupled to a Thermal Bath: Dynamics and Thermodynamics}

\begin{figure}
\centering
\subfigure[]{\label{IFT}
\includegraphics[scale=0.23]{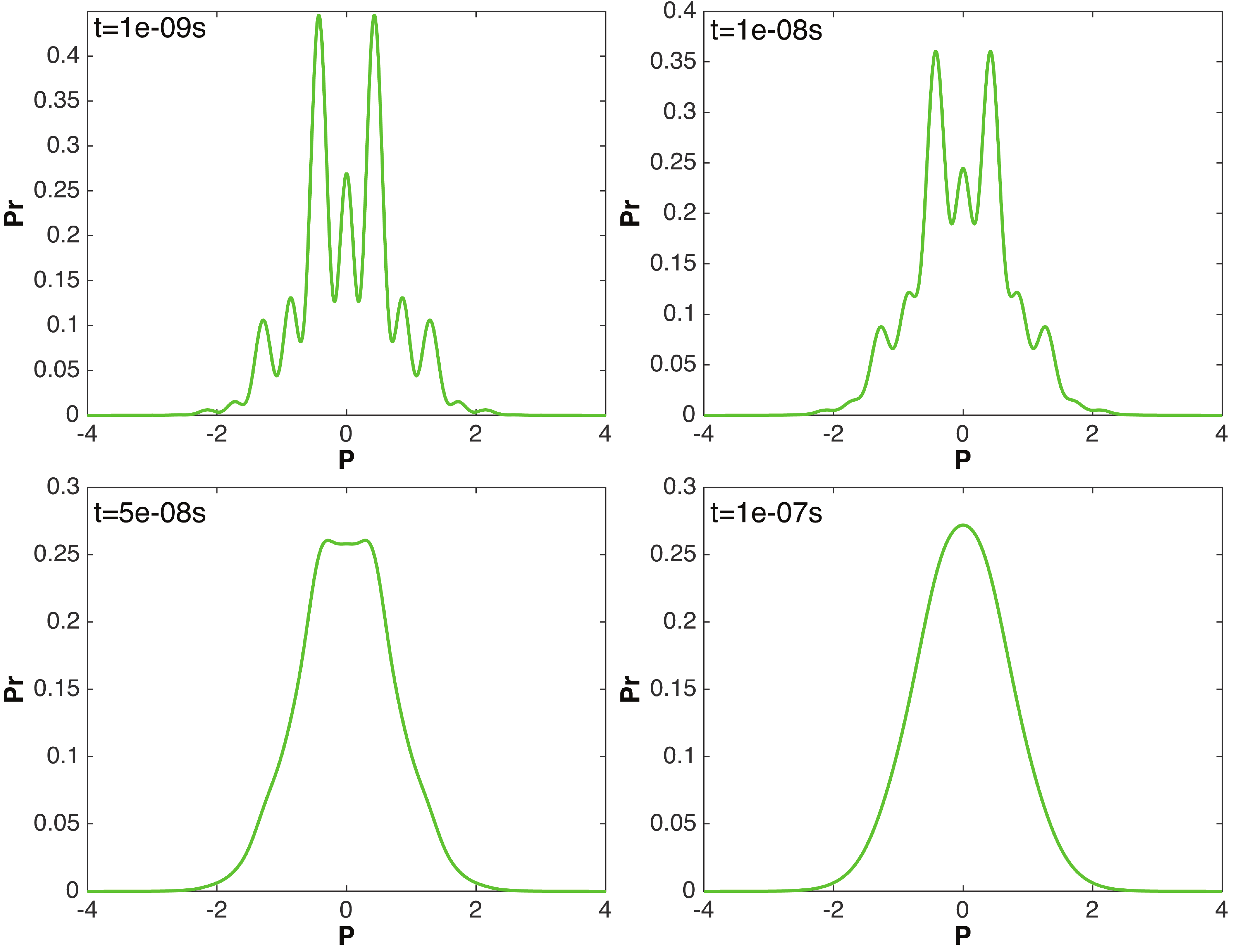}}
\subfigure[]{\label{IFadt}
\includegraphics[scale=0.4]{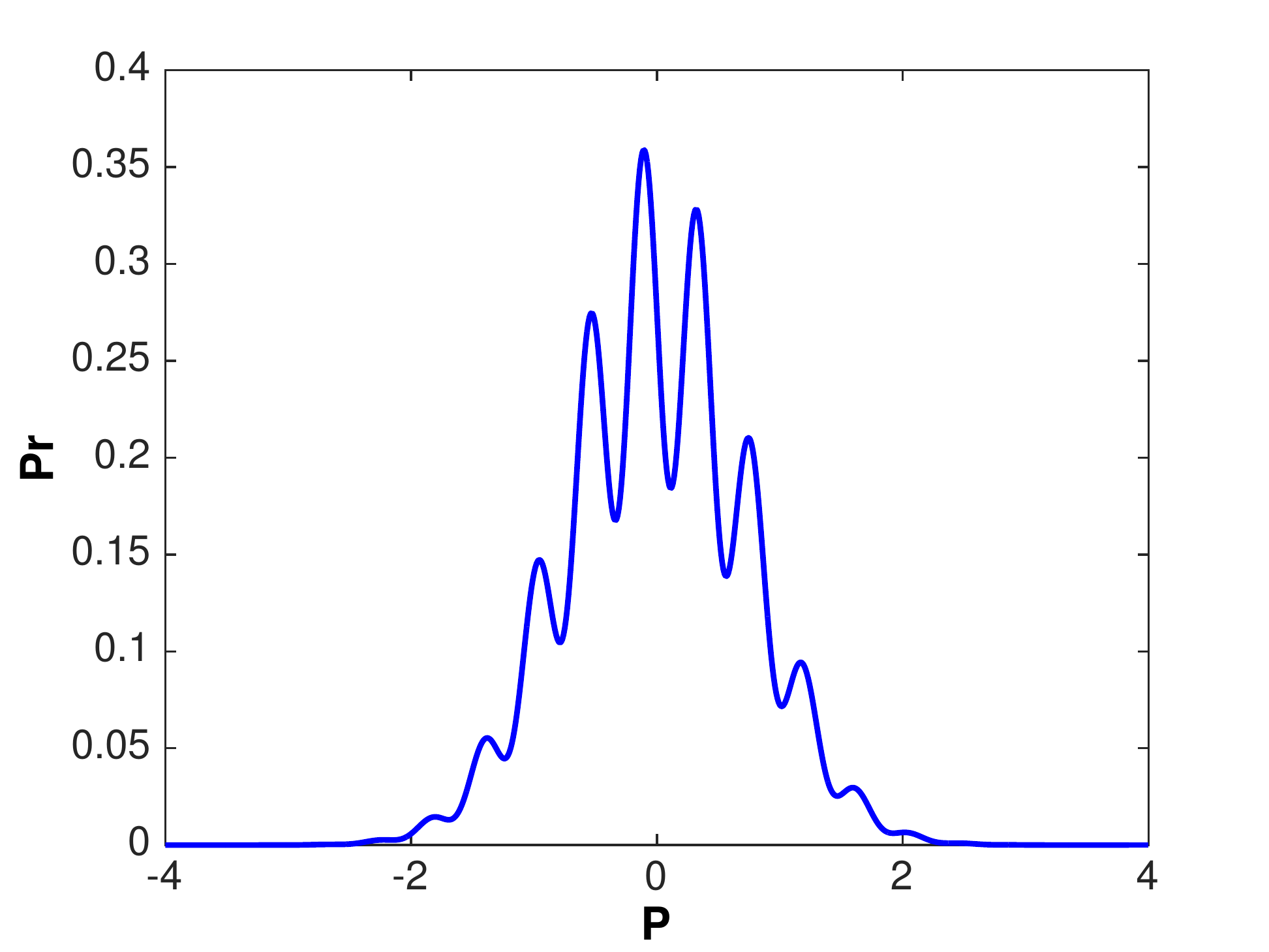}}
\caption{Interference patterns with $C_1=C_2=1/\sqrt{2}$ in (a) $\log(\frac{\Omega}{T}s.K)=9.52$ (room temperature) from $t=1\times 10^{-9}s$ to $t=1\times 10^{-7}s$, in which the interference pattern completely vanishes and (b) $\log(\frac{\Omega}{T}s.K)=23$ (low temperature) when the decoherence process is completed $(\eta=0)$.}
\label{IF}
\end{figure}

In this section, we study the dynamic and the thermodynamic properties of a harmonic oscillator entering an interferometer that is partially coupled to a thermal bath FIG. \ref{inter}. A harmonic oscillator in the state $\vert 1\rangle$ enters the interferometer through the beamsplitter $BS_1$.  As described below, the particle reflects with probability $\vert C_1\vert^2$, reaches the detector $D_1$, and/or goes through the $50$:$50$ beamsplitter $BS_2$  with probability $\vert C_2\vert^2=1-\vert C_1\vert^2$. The particle interacts with unitary gates $U_1$ and $U_2$ in paths (b) and (c), respectively. According to the system's Hamiltonian, the unitary operators $\hat{U}_1$ and $\hat{U}_2$ are defined below in a $4\text{D}$ Hilbert space to increase the particle energy
\begin{align}
\label{Us}
\hat{U}_1=
\begin{pmatrix}
1&0&0&0 \\
0&0&1&0 \\
0&1&0&0 \\
0&0&0&1
\end{pmatrix}
~~~
\hat{U}_2=
\begin{pmatrix}
1&0&0&0 \\
0&0&0&1 \\
0&0&1&0 \\
0&1&0&0
\end{pmatrix}.
\end{align}
Therefore, the state of the system just before the interaction with thermal baths is the following state ($a,b$ and $c$ refer to interferometer branches in FIG. \ref{inter})
\begin{align}
\label{istate}
C_1\vert 1\rangle_a\vert 0\rangle_b\vert 0\rangle_c +\frac{C_2}{\sqrt{2}}e^{i\phi}\vert 0\rangle_a\vert 2\rangle_b\vert 0\rangle_c+i\frac{C_2}{\sqrt{2}}\vert 0\rangle_a\vert 0\rangle_b\vert 3\rangle_c.
\end{align}
For the sake of simplicity, in \eqref{istate}, we substitute the states $\vert 1\rangle_a\vert 0\rangle_b\vert 0\rangle_c$, $\vert 0\rangle_a\vert 2\rangle_b\vert 0\rangle_c$ and $\vert 0\rangle_a\vert 0\rangle_b\vert 3\rangle_c$ with $\vert 0\rangle$, $\vert 1\rangle$ and $\vert 2\rangle$, respectively. This is just a mathematical simplification and does not affect the physical description of the problem. Thus, for \eqref{istate} one gets
\begin{align}
C_1\vert 0\rangle +\frac{C_2}{\sqrt{2}}e^{i\phi}\vert 1\rangle +i\frac{C_2}{\sqrt{2}}\vert 2\rangle.
\end{align}
We describe the system interaction with the environment in terms of decoherence theory, applying the master equation \eqref{ME}. According to \eqref{istate}, the density matrix of the system before its coupling with the thermal bath is
\begin{align}
\label{idm}
\begin{pmatrix}
\vert C_1\vert^2 & \frac{C_1C_2^*}{\sqrt{2}}e^{-i\phi} & -i\frac{C_1C_2^*}{\sqrt{2}} \\
\frac{C_1^*C_2}{\sqrt{2}}e^{i\phi} & \frac{\vert C_2\vert^2}{2} & -i\frac{\vert C_2\vert^2}{\sqrt{2}}e^{i\phi} \\
i\frac{C_1^*C_2}{\sqrt{2}} & i\frac{\vert C_2\vert^2}{\sqrt{2}}e^{-i\phi} & \frac{\vert C_2\vert^2}{2}
\end{pmatrix}.
\end{align}
For such a system, the master equation \eqref{ME} can be solved under the rotating-wave approximation. Due to the system interactions with thermal baths in the path $(b)$ and $(c)$, solving the master equation \eqref{ME}, considering the effect of the beamsplitter $BS_3$, results in the following density matrix
\begin{widetext}
\begin{equation}
\label{bdm}
\hat{\rho}_s(t)=
\begin{pmatrix}
\vert C_1\vert^2 & \frac{C_1C_2^*}{2}\eta^{1/2}(e^{-i\phi}-1) & -i\frac{C_1C_2^*}{2}\eta^{1/2}(e^{-i\phi}+1) \\
\frac{C_1^*C_2}{2}\eta^{1/2}(e^{i\phi}-1) & \frac{\vert C_2\vert^2}{2}(1-\eta\cos\phi) & i\vert C_2\vert^2(\frac{\eta^2-1}{2(2\bar{n}+1)})+\frac{\vert C_2\vert^2}{2}\eta\sin\phi \\
i\frac{C_1^*C_2}{2}\eta^{1/2}(e^{i\phi}+1) & -i\vert C_2\vert^2(\frac{\eta^2-1}{2(2\bar{n}+1)})+\frac{\vert C_2\vert^2}{2}\eta\sin\phi & \frac{\vert C_2\vert^2}{2}(1+\eta\cos\phi)
\end{pmatrix},
\end{equation}
\end{widetext}
where $\eta=e^{-\Gamma t(2\bar{n}+1)}$. The probability distribution of the off-diagonal elements of $\hat{\rho}_s(t)$ can be calculated by $\text{Pr}(P)=\langle P\vert \hat{\rho}_s(t)\vert P\rangle$, which represents the interference fringes along the $P$ axis. Using the well-known harmonic oscillator wave functions, for the interference fringes, one gets 
\begin{align}
\text{Pr}(P)=&\sqrt{\frac{1}{\Omega\pi}}e^{-\frac{P^2}{\Omega}}\bigg\{1+Z\eta^{\frac{1}{2}}\big(\cos(\frac{Pd}{2})\big[\sin\theta-\cos\theta \nonumber \\
&+\cos(\phi-\theta)-\sin(\phi-\theta)\big]+\sin(\frac{Pd}{2})\big[\sin\theta  \nonumber \\
&-\cos\theta+\sin(\phi-\theta)-\cos(\phi-\theta)\big]\big)+\vert C_2\vert^2 \nonumber \\
&\times\big[\eta\sin\phi\cos(Pd)+\frac{\eta^2-1}{2\bar{n}+1}\sin(Pd)\big] \bigg\},
\end{align}
where $C_1C_2^*=Ze^{i\theta}$ and $d$ represents the differences in optical path lengths. At high temperatures, the terms $\eta\sin\phi\cos(Pd)$ and $Z\eta^{1/2}(\cdots)$ are responsible for the interference pattern, as is illustrated in FIG. \ref{IFT}. With the passage of time, the interference vanishes due to decoherence process. At the low-temperature limit, however, the term $\frac{\eta^2-1}{2\bar{n}+1}\sin(Pd)$ is at work, albeit when decoherence process is completed and  $\eta\rightarrow 0$ (FIG. \ref{IFadt}).

To study the Clausius inequality within the system-baths interaction process, we employ the approach introduced by Nieuwenhuizen and Allahvedyan \cite{Nie,All2}. In this framework, the expectation value of the energy of the system is considered as its internal energy 
\begin{equation}
\label{U}
U=\langle \hat{H}_s\rangle=\frac{\langle \hat{P}^2\rangle}{2}+\frac{\Omega^2\langle \hat{X}^2\rangle}{2}.
\end{equation}
Moreover, a quasistatic variation of $U$ is
\begin{eqnarray}
\label{dU}
\text{d}U=&&\int\int \text{d}\mathcal{X}\text{d}\mathcal{P}W(\mathcal{X},\mathcal{P})\text{d}\hat{H}_s(\mathcal{X},\mathcal{P}) \nonumber \\
&+&\int\int \text{d}\mathcal{X}\text{d}\mathcal{P}\hat{H}_s(\mathcal{X},\mathcal{P})\text{d}W(\mathcal{X},\mathcal{P}),
\end{eqnarray}
where $W(\mathcal{X},\mathcal{P})$ represents the Wigner distribution function of the system in the position-momentum space. The first term in the right-hand side of Eq. \eqref{dU} is known as the work done on the system  $\delta W$; which is zero in our case because of the Hamiltonian that is independent of time. The second term is also the heat exchanged with the bath $\delta Q$. Regarding the zero work transfer, via Eq. \eqref{U} for the exchanged heat, one gets
\begin{equation}
\label{Q}
\delta Q=\left( \frac{\Omega^2}{2}\frac{\partial\langle \hat{X}^2\rangle}{\partial t}+\frac{1}{2}\frac{\partial\langle \hat{P}^2\rangle}{\partial t} \right) \text{d}t.
\end{equation}
Now, for calculating the heat, the position and the momentum quadratures ($\langle \hat{X}^2\rangle$ and $\langle \hat{P}^2\rangle$) are needed. Again, by considering $C_1C_2^*=Ze^{i\theta}$, with respect to the density matrix \eqref{bdm} in the Ohmic regime, one obtains
\begin{align}
\langle \hat{X}^2\rangle=\frac{1}{2\Omega}[1&+\frac{\vert C_2\vert^2}{2}(3-\eta\cos\phi) \nonumber \\
&+\sqrt{2}Z\eta^{\frac{1}{2}}(\sin\theta-\sin(\phi-\theta))] \\
\langle \hat{P}^2\rangle=\frac{\Omega}{2}[1&+\frac{\vert C_2\vert^2}{2}(3-\eta\cos\phi) \nonumber \\
&+\sqrt{2}Z\eta^{\frac{1}{2}}(\sin(\phi-\theta)-\sin\theta)].
\end{align}
Consequently, by integrating Eq. \eqref{Q}, one can calculate the exchanged heat as
\begin{equation}
\label{heat}
Q(t)=\frac{\Omega\cos\phi}{4}\vert C_2\vert^2(1-\eta),
\end {equation}
which is time-dependent equation and varies with $\vert C_2\vert^2$. As we expect, with the passage of time, more heat is exchanged during the process. In this regard, Eq. \eqref{heat} shows that the transparency of beamsplitter $BS_1$ causes an increase in the exchanged heat. It has to be noted that this is not the case for $\vert C_2\vert^2=1$, in which the system is studied in a 2D Hilbert space. In such a situation, the position and the momentum quadratures are independent of time. Thus, there is no heat transfer in the process, and the process is in agreement with the Clausius inequality \cite{Sol}.

In the quantum regime, we use the von Neumann entropy $S(\hat{\rho})=-\text{Tr}[\hat{\rho}\log(\hat{\rho})]$ as the thermodynamic entropy \cite{All2,Hil}. Generally, the von Neumann entropy has no relation with thermodynamic entropy. However, it is the best choice we have and the only definition that fulfills the properties of an entropy function (such as additivity and subadditivity) \cite{Lie,Wehr}. Moreover, due to decoherence theory, we expect that the system rests in a thermal state (completely mixed state) when the decoherence process is completed. The von Neumann entropy is the only definition that is identical to the thermodynamic entropy in thermal states \cite{Bre}. Since the initial entropy $S_0$ is zero, we have $\Delta S=S_t$. To our expectation, for the Lindblad process, the von Neumann entropy increases with time. Therefore, positive values of the following relation 
\begin{equation}
\label{F}
F(t):=S_t-\frac{Q(t)}{T},
\end{equation}
 satisfies the Clausius inequality.
 
In the next section, we discuss the conditions in which the system behavior is not consistent with the Clausius Inequality.

\section{Results and Discussion}

\begin{figure}[!t]
\centering
\subfigure[]{\label{violation}
\includegraphics[scale=0.45]{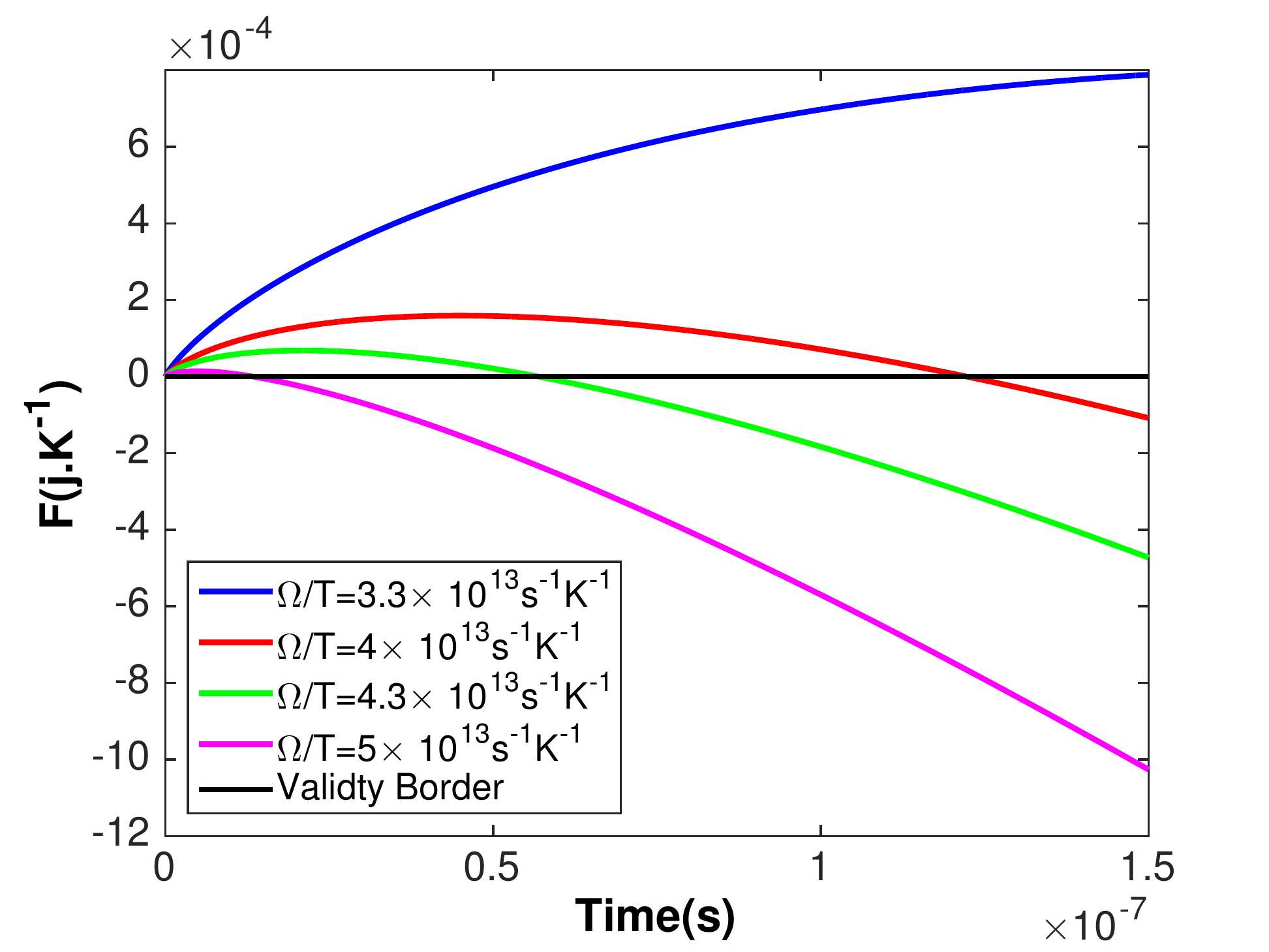}}
\subfigure[]{\label{Fdensity}
\includegraphics[scale=0.45]{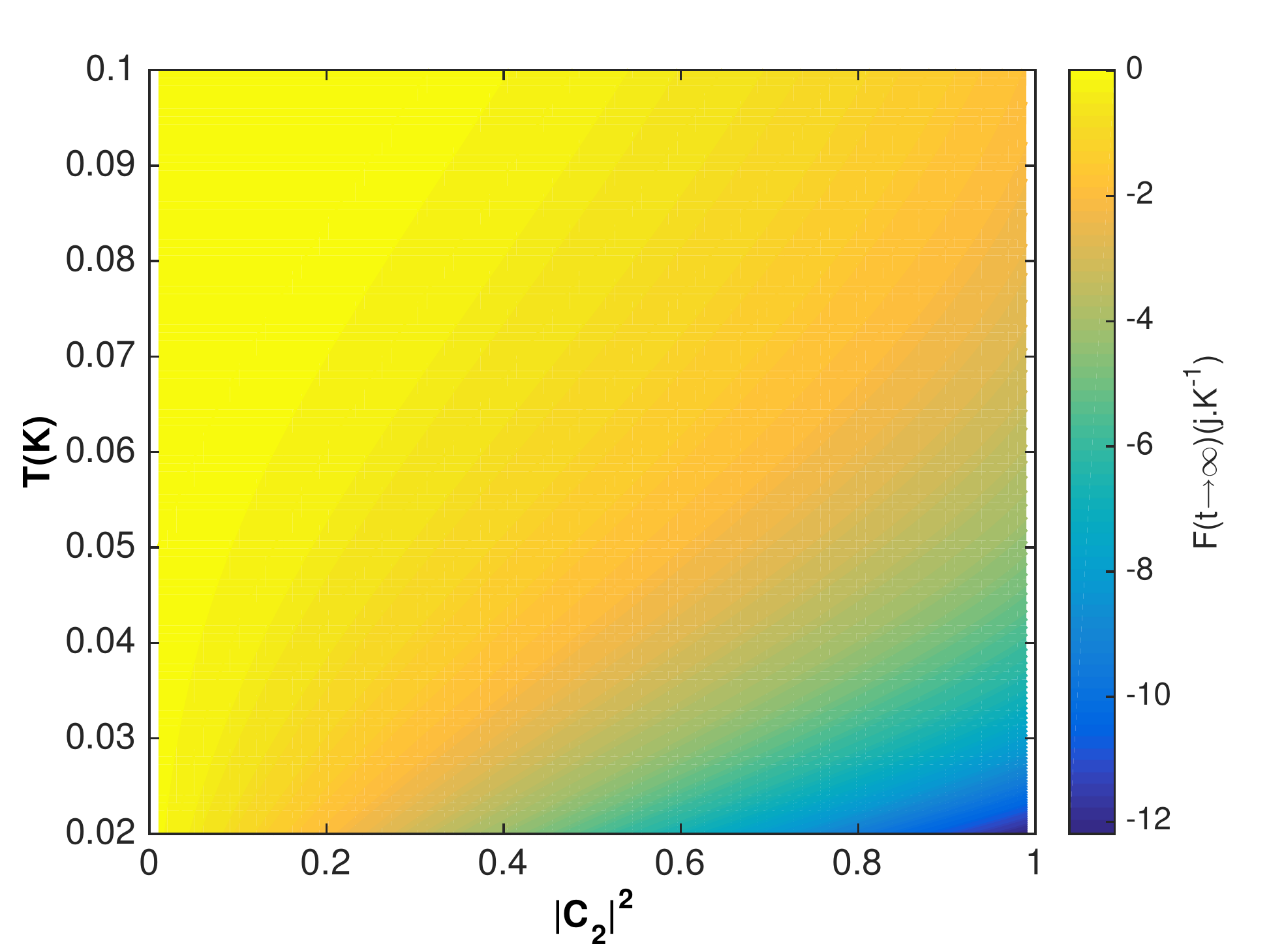}}
\caption{(a) The regions where the system violates the Clausius inequality versus time for different values of $\Omega/T$ and $C_1=C_2=1/\sqrt{2}$. Positive $F$ ensures us that the system satisfies the inequality. Higher amounts of $\Omega/T$ cause more violation from the Clausius inequality. (b) The violation of Clausius inequality at different temperatures and $\vert C_2\vert^2$ with $\Omega=1\times 10^{12}s^{-1}$, while the decoherence process is complete. Low temperatures and more transparency of the beamsplitter $BS_1$ cause stronger violations. The values of $\vert C_2\vert^2=0$ and $1$, are not included, due to the dimensions of the Hilbert space here.}
\label{IF}
\end{figure}

The Clausius inequality in Lindblad regime is typically validated with an increase in the entropy with time. However, the violation of the Clausius inequality has been reported in some situations, especially when the system-environment are entangled \cite{All2}. It has been argued that although the effect of the system-environment entanglement vanishes exceeding some critical temperatures, the violation from Clausius inequality is expected yet \cite{Hil}. Here, as is apparent in FIG. \ref{violation}, with the passage of time, the Clausius inequality is violated at low temperatures or precisely high values of  $\Omega/T$. On the other hand, no violation is observed at a high-temperature limit.

After the decoherence process ($t\rightarrow\infty$), the system entropy is maximum
\begin{align}
\label{entropy-changes}
S_\infty=&-\vert C_2\vert^2(\frac{\bar{n}}{2\bar{n}+1}\log\frac{\bar{n}}{2\bar{n}+1}+\frac{\bar{n}+1}{2\bar{n}+1}\log\frac{\bar{n}+1}{2\bar{n}+1}) \nonumber \\
&-\vert C_1\vert^2\log\vert C_1\vert^2-\vert C_2\vert^2\log\vert C_2\vert^2.
\end{align}
As FIG. \ref{violation} shows, at lower temperatures, although the entropy increases with time, it exceeds the boundary of Clausius inequality (i.e., the lines with negative $F$). Interestingly, since the Born approximation is assumed during the derivation of the master equation \eqref{ME}, the system-environment state remains in an approximate product state for all times $\hat{\rho}(t)\approx\hat{\rho}_s(t)\otimes\rho_\varepsilon$ \cite{Schl}. This state can be fulfilled because, according to the Born assumption, the environment is quite large and its state remains nearly invariant. Therefore, both the system and the environment remain separable. Also, the strength of the system-environment coupling has no effect, since both the entropy \eqref{entropy-changes} and heat term \eqref{heat} are independent of $\gamma_0$. Accordingly, the observed violation is not an outcome of the system-environment entanglement. Moreover, regarding FIG. \ref{Fdensity}, the beamsplitter $BS_2$ transparency also plays a role in the amount of the entropy violation. It is of note that the mathematical representation of the entropy is pretty arbitrary. The standard definition of the entropy is the “number of microstates”, which is interpreted as the number of pure states in quantum mechanics. As the entropy increases in a Lindblad regime, the system evolves to a mixed state, independent of the mathematical definition of the entropy. Therefore, the physical behavior of the system would be the same as reported here. In other words, any alternative representation of thermodynamic entropy would affect the quantities not the evolution of the system.

FIG. \ref{IFadt} presents the possibility of the system to remain coherent at the low-temperature limit (we studied this issue in \cite{Sol}). Moreover, the entropy of the system exceeds the boundary of the Clausius inequality at low temperatures (FIG. \ref{Fdensity}). In this regard, we need to study the progression of the coherence of the system state through the interaction of the system with the thermal bath. The first axiomatic approach in quantifying coherence was introduced by Aberg and developed by Baumgratz et al. \cite{Abe,Bau}. According to this axiomatic program, any quantifier of coherence must be consistent with the following postulates: {\it nonnegativity} and {\it Monotonicity} (i.e., the reduction under the action of incoherent operations).  Moreover, if the coherence quantifier fulfills all the postulates mentioned in TABLE \ref{table1}, we call it a {\it coherence measure} \cite{Str}. In the current literature, the most relevant coherence quantifiers are distillable coherence \cite{Win}, distance-based coherence quantifiers \cite{Bau}, convex root coherence quantifiers \cite{Yua} and coherence monotones from entanglement \cite{Str2}. From the stated categories, only the distillable coherence and the relative entropy (as a distance-based coherence quantifier) fulfill all the postulates and are considered as the coherence measure. Besides, it has been proven that the relative entropy coherence measure is equal to the distillable coherence \cite{Str}.

\begin{table}[!t]
\centering
\caption{Requirements for a coherence measure $C(\hat{\varrho})$. $S$ is von Neumann entropy and $\hat{\Xi}$ represents the dephasing operator.  \label{table1}}
\begin{tabular}{p{17ex} | p{30ex}}
\hline
Postulate & Definition \\
\hline
Nonnegativity & $C(\hat{\varrho})\geq 0$. \\[1ex]
Monotonicity & {\small $C$ does not increase under the action of incoherent operations.} \\[1ex]
Strong monotonicity & {\small $C$ does not increase on average under selective incoherent operations.} \\[1ex]
Convexity & {\small $C$ is a convex function of the state.}  \\[1ex]
Uniqueness  & {\small For any pure states $\vert\psi\rangle$, $C$ takes the form: \linebreak $C(\vert\psi\rangle\langle\psi\vert)=S(\hat{\Xi}[\vert\psi\rangle\langle\psi\vert])$.} \\[1ex]
Additivity & {\small $C$ is additive under tensor products.}  \\[1ex]
\hline
\end{tabular}
\end{table}

\begin{figure}
\centering
\includegraphics[scale=0.45]{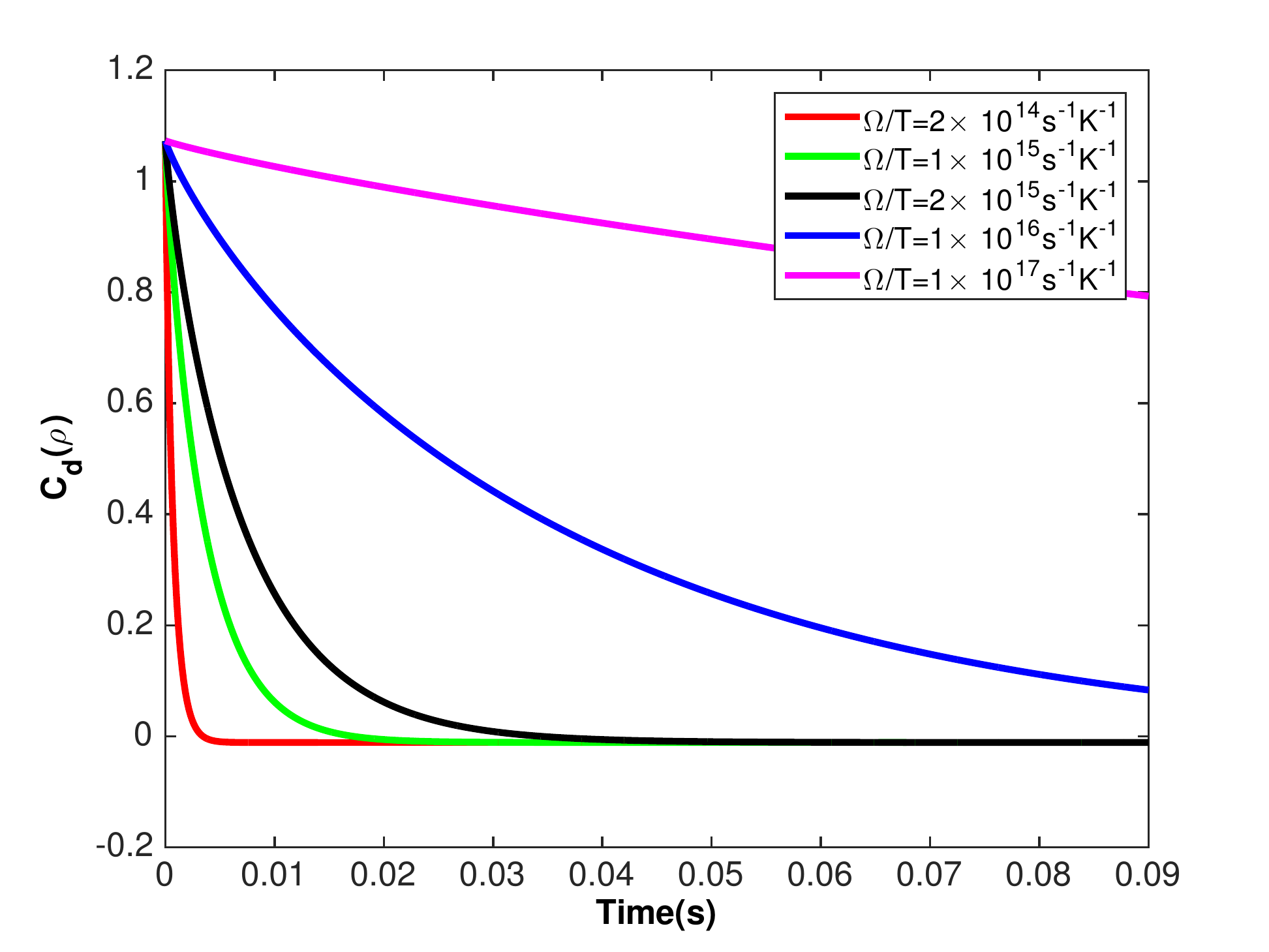}
\caption{The distillable coherence $C_d(\hat{\rho})$ versus time in a variety of $\Omega/T$ values with $\vert C_2\vert^2=0.6$. The coherence of the system decreases with a more slight slope at lower temperatures.}
\label{coherence}
\end{figure}

The distillable coherence is the optimum number of maximally coherent states that can be obtained from a state $\hat{\varrho}$ using incoherent operations. One can write the distillable coherence as \cite{Str,Win}
\begin{equation}
\label{DistillableCoherence}
C_d(\hat{\varrho})=S(\hat{\Xi}[\hat{\varrho}])-S(\hat{\varrho}),
\end{equation}
where $\hat{\Xi}[\hat{\varrho}]=\sum_{i=0}^{d-1}\vert i\rangle\langle i\vert\hat{\varrho}\vert i\rangle\langle i\vert$ is the dephasing operator. The distillable coherence $C_d(\hat{\rho})$ is plotted for different values of $\Omega/T$ in FIG. \ref{coherence}. As can be seen, the coherence decreases with a slight slope at low temperatures, suggesting that the quantum dynamics necessitates the system to remain coherent (FIG. \ref{IFadt}), at the price of the contradiction with the Clausius inequality.

Regarding FIGs. \ref{Fdensity} and \ref{coherence}, the system after the interaction with thermal baths, tends to behave classically except for the low-temperature condition in which it keeps its quantum properties. Moreover, the figures show that the process is not consistent with the Clausius inequality. It is of note that, in this model, wherein the system and the environment are separable (Born approximation), the violation of the Clausius inequality is not directly resulted from the system-environment entanglement \cite{All2}. In order to investigate this issue in a better way, we review the case of work extraction from a coherent state in a cyclic unitary evolution \cite{Bin,All3,Pus}. In a cyclic unitary work extraction, the system with non-passive density matrix 
\begin{equation}
\label{rho}
\hat{\rho}=\sum r_n\vert r_n\rangle\langle r_n\vert~~~\text{with}~~~r_{n+1}\leq r_n \forall n,
\end{equation}
ends up in a so-called passive state $\hat{\pi}$, respecting the system Hamiltonian
\begin{equation}
\label{Hp}
\hat{H}=\sum \varepsilon_n\vert \varepsilon_n\rangle\langle \varepsilon_n\vert~~~\text{with}~~~\varepsilon_{n+1}\geq \varepsilon_n \forall n.
\end{equation}
Passive states are diagonal in the eigenbasis of Hamiltonian with decreasing populations for increasing energy levels
\begin{equation}
\label{pi}
\hat{\pi}=\sum \varepsilon_n\vert r_n\rangle\langle r_n\vert~~~\text{with}~~~\varepsilon_{n+1}\leq \varepsilon_n \forall n.
\end{equation}
Passive states are in agreement with the second law of thermodynamics in Kelvin-Planck formulation. This statement of the second law of thermodynamics expresses that it is impossible to devise a cyclically operating device that absorbs energy in the form of heat from a single thermal reservoir and delivers an equivalent amount of work. Passive states can produce no work in a Hamiltonian cyclic process, where system returns to its initial Hamiltonian. Such a process can be described by a unitary operation $U$. Hence, the maximum extractable work from the system is
\begin{align}
W_\text{max}(\hat{\rho})=\operatorname*{max}_U \text{tr}\left[ H\left(\rho-U\rho U^{\dagger} \right) \right],
\end{align}
which is zero for passive states. It has to be noted that Gibbs states are consequently passive states \cite{Per}. The process of unitary cyclic work extraction from a non-passive state \eqref{rho}, with respect to the Hamiltonian \eqref{Hp} is called ergotropy \cite{Bin,Pus}:
\begin{equation}
\label{ergotropy}
\mathcal{W}=\text{tr}[\hat{\rho} \hat{H}-\hat{\pi} \hat{H}]=\sum_{m,n}r_m\varepsilon_n[\vert\langle\varepsilon_n\vert r_m\rangle\vert^2-\delta_{mn}],
\end{equation}
which is equivalent to the quantity $W_\text{max}$.

\begin{figure}[!t]
\centering
\includegraphics[scale=0.45]{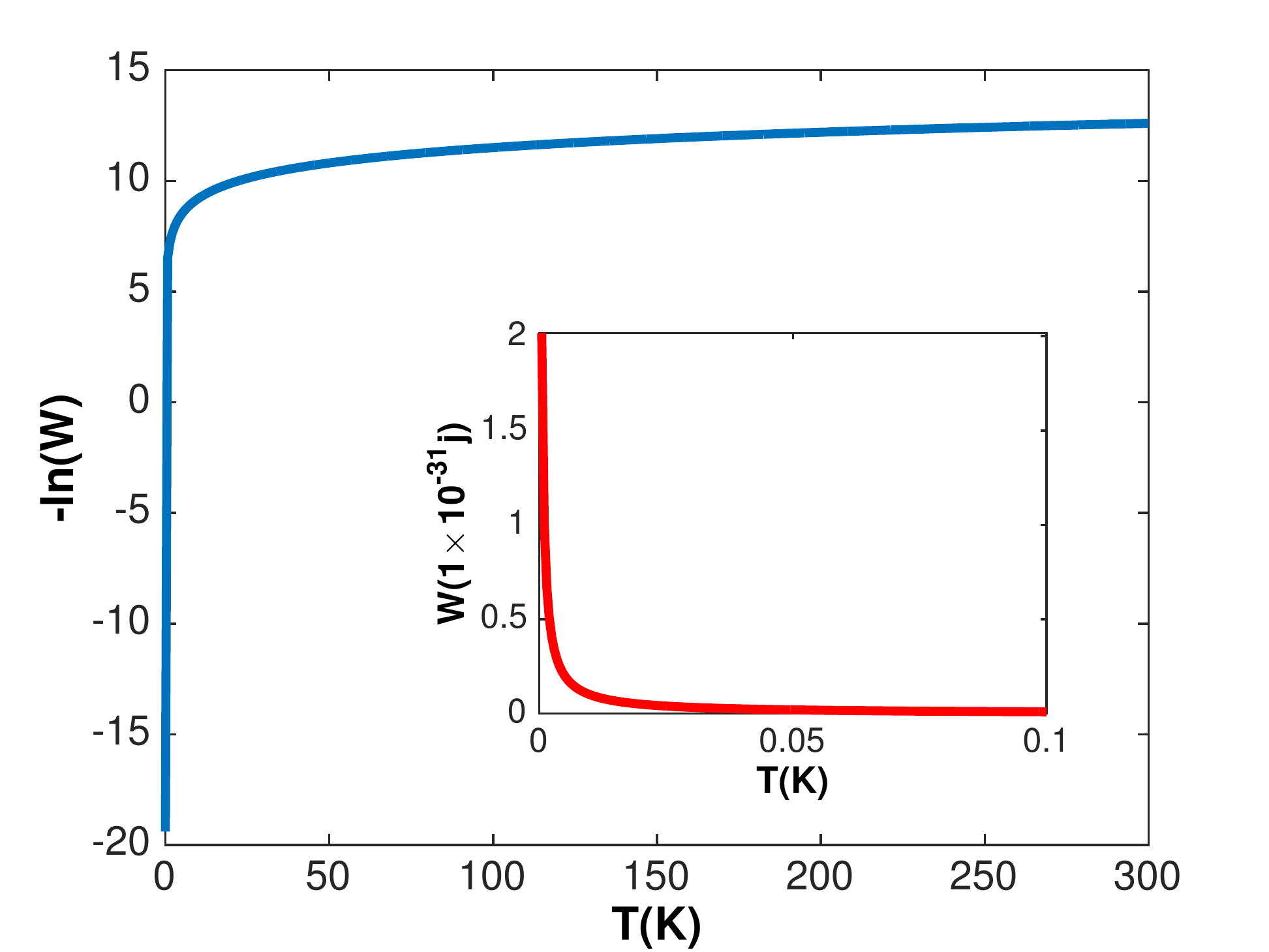}
\caption{Blue Graph: The changes of $-\text{ln}(\mathcal{W}\times10^{31})$ versus the temperature evolution from $T=0K$ to the room temperature. Red Graph: The behavior of ergotropy ($\mathcal{W}$) at low temperatures.}
\label{ergofig}
\end{figure}

In our case, the system's state \eqref{bdm}, after the decoherence process ($t\rightarrow\infty$), ends up in the following non-passive state:
\begin{align}
\label{nps}
\hat{\rho}_s(\infty)=
\begin{pmatrix}
\vert C_1\vert^2 & 0 & 0 \\
0 & \frac{\vert C_2\vert^2}{2} & \frac{-i\vert C_2\vert^2}{2(2\bar{n}+1)} \\
0 & \frac{i\vert C_2\vert^2}{2(2\bar{n}+1)} & \frac{\vert C_2\vert^2}{2} 
\end{pmatrix}.
\end{align}
We can extract work (ergotropy) from \eqref{nps} at the price of coherence and leave the system in the resulting passive state
\begin{align}
\label{pis}
\hat{\pi}_s=
\begin{pmatrix}
\vert C_1\vert^2 & 0 & 0 \\
0 & \frac{\vert C_2\vert^2}{2}+\frac{\vert C_2\vert^2}{2(2\bar{n}+1)} & 0 \\
0 & 0 & \frac{\vert C_2\vert^2}{2}-\frac{\vert C_2\vert^2}{2(2\bar{n}+1)}
\end{pmatrix}.
\end{align}
Obviously, for \eqref{pis}, we have $C_d(\hat{\pi}_s)=0$. Therefore, with the price of coherence, we extract work from the system and reach a state that is compatible with the Second Law of Thermodynamics. According to Eq. \eqref{ergotropy}, the maximum work extraction (ergotropy) from \eqref{nps}, respecting the Hamiltonian $\hat{H_s}$, is equal to
\begin{equation}
\label{ers}
\mathcal{W}_s=\frac{\vert C_2\vert^2}{2(2\bar{n}+1)}\hbar\Omega.
\end{equation}
As Eq. \eqref{ers} shows and we plotted in FIG. \ref{ergofig}, ergotropy tends to zero with an increase in temperature. At high temperatures, the system coherence is nearly zero (Fig. \ref{coherence}). Subsequently, the system itself is in a passive state and work extraction is not possible (the ergotropy is zero). Therefore, the system's entropy obeys Clausius inequality. In comparison, at low temperatures, the system's entropy trespasses the Clausius inequality because, by non-zero unitary cyclic extraction, the system loses its coherence and ends up in the passive state. In other words, as the system’s coherence act as a source for work extraction, at the same time, the loss of coherence causes the system to satisfy the Clausius inequality. Therefore, the resulted agreement with the Clausius inequality from the loss of the system coherence confirms that the quantum coherence is the main reason for system violating the Clausius inequality.

\section{conclusion}
The main objective of the current study is to propose a model to challenge our understandings of thermodynamic concepts in the quantum regime. For this purpose, we investigated the quantum approach to Clausius inequality in an interferometer that partially is coupled to thermal baths. Interestingly, our observations, at first sight, was in contradiction with our expectations; accordingly, we sought for reasons for such behaviors. 

We studied a harmonic oscillator entering an interferometer. Later, the system resulted in a superposition of the particle path in three branches of the interferometer. In the first branch, the particle directly reached the detector but in the other two, the particle interacted with thermal baths. In this approach, the first branch that has no interaction with thermal baths makes the system to resist against losing its coherence. By illustrating the interference pattern at high and low temperatures, we observed that despite the effect of decoherence process, the system remains coherent in the low-temperature limit. To our astonishment, the system entropy exceeded the Clausius inequality in this region. We discussed this observation as a matter of system coherence. Consequently, plotting the distillable coherence in a variety of temperatures revealed that when the system is kept coherent, it is observed that the Clausius inequality is violated. 

In a search for the origin of the stated violation, we referred to the unitary cyclic work extraction process (i.e., ergotropy). By extracting work, the system loses its coherence and ends up in a passive state that is diagonal in the energy eigenbasis and obeys the Kelvin-Planck formulation of the Second Law of Thermodynamics. As temperature declines, an increase occurs in the amount of ergotropy. We argued that at high temperatures in which the system itself is incoherent and the ergotropy is nearly zero, the system obeys the Clausius inequality. On the other hand, at low temperatures and the non-zero ergotropy, at the cost of losing coherence, the system becomes compatible with the Second Law of Thermodynamics. 

The violation from Clausius inequality was formerly reported in previous studies as a matter of the system-environment entanglement. Later, the phenomenon was observed also in conditions with no entanglement. Accordingly, the reason for such behavior remained open to discussion until now. In our model, we observed the system violation from Clausius inequality just in the low-temperature range. The calculated thermodynamic quantities (such as entropy and heat) were independent of the system-bath coupling and also the system-bath entanglement was assumed to be nearly zero (Born approximation). The only terms that vary in the system final state, from high-temperature to low-temperature limit, are the non-diagonal elements of the density matrix. The non-diagonal terms tend to zero as the temperature of the baths increases. Also, with zero non-diagonal terms, no violation was observed. The direct relation of the system's coherence with the non-diagonal terms of the density matrix led us considering the coherence as the reason for the mentioned violation from the Clausius inequality. As we expected, at low temperatures, the system is more coherent and also violates the Clausius inequality. Moreover, by extracting system's coherence in the form of work (ergotropy), the system ended up in a passive state and, subsequently, no violation was observed. Coherence is the main characteristic of quantum nature. The system’s coherence causing the violation from Clausius inequality indeed shows that the Clausius inequality is incompatible with the quantum nature of the system. With further investigation on the problem, we found out that this observation is not entirely unexpected. According to information theory and the concept of entropy, the Clausius inequality is an upper bound for the information storage capacity. This study shows that due to the coherence in quantum systems, theoretically, we are capable of storing more information in quantum states rather than the classical ones. Subsequently, losing coherence as a consequence of interaction with the environment causes the loss of information. In this regard, further work needs to be done to investigate the upper bound for information storage in the quantum systems. Besides, since the developments in the quantum thermal machines, such studies find special importance because of less limitations for the thermodynamic processes. Hence, we potentially have external work extractions from quantum systems, more efficiency in quantum thermal machines, and more information storage in quantum systems.

\section*{Acknowledgments}
We thank Dr. Farhad Taher Ghahramani and Dr. Hamidreza Naeij for their constructive comments on the presentation of the paper. 



\begin{thebibliography}{1}
\bibitem{Gold} M. Goldstein and F. I. Goldstein, {\it The Refrigerator and the Universe}, Harvard University Press, Cambridge (1993).
\bibitem{Liu} N. Liu, J. Goold, I. Fuentes, V. Vedral, K. Modi, D.E. Bruschi, Class. Quantum Grav. {\bf 33}, 035003 (2016).
\bibitem{Gem} J. Gemmer, M. Michel and G. Mahler, {\it Quantum Thermodynamics: Emergence of Thermodynamic Behavior Within Composite Quantum Systems}, Lect. Notes Phys. 784, Springer, Berlin Heidelberg (2009).
\bibitem{Goo1} J. Goold, M. Huber, A. Riera, L. del Rio and P. Skrzypczyk, J. Phys. A: Math. and Theo. {\bf 49}, 143001 (2016).
\bibitem{Vin} S. Vinjanampathy and J. Anders, Cont. Phys. {\bf 57}, 545-579 (2016).
\bibitem{Bra} F. G. Brandao, M. Horodecki, J. Oppenheim, J. M. Renes and R. W. Spekkens, Phys. Rev. Lett. {\bf 111}, 250404 (2013).
\bibitem{Bin} F. Binder, S. Vinjanampathy, K. Modi and J. Goold, Phys. Rev. E {\bf 91}, 032119 (2015).
\bibitem{Deu} J. Deutsch, New J. Phys. {\bf 12}, 075021 (2010).
\bibitem{Cwi} P. Ćwikliński, M. Studziński, M. Horodecki, J. Oppenheim, Phys. rev. lett. {\bf 115}, 210403 (2015).
\bibitem{Pop} S. Popescu, A. J. Short and A. Winter, Nat. Phys. {\bf 2}, 754-758 (2006).
\bibitem{del} L. del Rio, J. Aberg, R. Renner, O. Dahlsten and V. Vedral, Nature {\bf 474}, 61-63 (2011).
\bibitem{And} J. Anders and V. Giovannetti, New J. of Phys. {\bf 15}, 033022 (2013).
\bibitem{Kie} T. D. Kieu, Phys. rev. lett. {\bf 93}, 140403 (2004).
\bibitem{All} A. Allahverdyan and T. M. Nieuwenhuizen, Phys. Rev. E {\bf 71}, 066102 (2005).
\bibitem{Esp} M. Esposito and S. Mukamel, Phys. Rev. E {\bf 73}, 046129 (2006).
\bibitem{Land} L. D. Landau and E. M. Lifshitz, {\it Statistical Physics, Part 1}, Pergamon, London (1980).
\bibitem{All2} A. E. Allahverdyan and T. M. Nieuwenhuizen, Phys. Rev. Lett. {\bf 85}, 1799 (2000).
\bibitem{Kli} Yu. L. Klimontovich, {\it Statistical Theory of Open Systems}, Kluwer, Amsterdam (1997).
\bibitem{Eva} D .J. Evans, E. Cohen and G. Morriss, Phys. Rev. Lett. {\bf 71}, 2401 (1993).
\bibitem{Sag} T. Sagawa and M. Ueda, Phys. rev. lett. {\bf 100}, 080403 (2008).
\bibitem{Jac} K. Jacobs, Phys. Rev. E {\bf 86}, 040106 (2012).
\bibitem{Groo} S. R. De Groot and P. Mazur, {\it Non-equilibrium thermodynamics}, Courier Corporation (2013).
\bibitem{Gog} C. Gogolin and J. Eisert, {\it Equilibration, thermalisation, and the emergence of statistical mechanics in closed quantum systems}, Rep. Prog. Phys. {\bf 79}, 056001 (2016).
\bibitem{Nie} T. M. Nieuwenhuizen and A. E. Allahverdyan, Phys. Rev. E {\bf 66}, 036102 (2002).
\bibitem{Hen} M. J. Henrich, G. Mahler and M. Michel, Phys. Rev. E {\bf 75}, 051118 (2007).
\bibitem{Wei} H. Weimer, M. J. Henrich, F. Rempp, H. Schröder and G. Mahler, Europhys. Lett. {\bf 83}, 30008 (2008).
\bibitem{Esp2} M. Esposito, K. Lindenberg and C. Van den Broeck, New J. Phys. {\bf 12}, 013013 (2010).
\bibitem{Esp3} M. Esposito and C. Van den Broeck, Europhys. Lett. {\bf 95}, 40004 (2011).
\bibitem{Gem2} J. Gemmer and R. Steinigeweg, Phys. Rev. E {\bf 89}, 042113 (2014).
\bibitem{Wil} H. Wilming, R. Gallego and J. Eisert, Phys. Rev. E {\bf 93}, 042126 (2016).
\bibitem{San} J. P. Santos, G. T. Landi and M. Paternostro, Phys. Rev. Lett. {\bf 118}, 220601 (2017).
\bibitem{Deu2} J. M. Deutsch, Phys. Rev. A {\bf 43}, 2046 (1991).
\bibitem{Rig} M. Rigol, V. Dunjko and M. Olshanii, Nature {\bf 452}, 854 (2008).
\bibitem{Cov} T. M. Cover and J. A. Thomas, {\it Elements of information theory}, John Wiley and sons (2006).
\bibitem{Lin} N. Linden, S. Popescu, A. J. Short and A. Winter, Phys. Rev. E {\bf 79}, 061103 (2009).
\bibitem{Sho} A. J. Short and T. C. Farrelly, New J. Phys. {\bf 14}, 013063 (2012).
\bibitem{Rie} A. Riera, C. Gogolin and J. Eisert, Phys. Rev. Lett. {\bf 108}, 08040 (2012).
\bibitem{Hor1} M. Horodecki and J. Oppenheim, Int. J.Mod. Phys. {\bf 27}, 1345019 (2012).
\bibitem{Ple} M.B. Plenio and V. Vitelli, Contemp. Phys. {\bf 42}, 25-60 (2001).
\bibitem{Par} J.M. Parrondo, J.M. Horowitz and T. Sagawa, Nat. Phys. {\bf 11}, 131-139 (2015).
\bibitem{Hil} S. Hilt and E. Lutz, Phys. Rev. A {\bf 79}, 010101 (2009).
\bibitem{Hor} C. H\"orhammer and H. B\"uttner, J. Stat. Phys. {\bf 133}, 1161-1174 (2008).
\bibitem{Lost} M. Lostaglio, D. Jennings and T. Rudolph, Nat. commun. {\bf 6}, 6383 (2015).
\bibitem{Lost2} M. Lostaglio, K. Korzekwa, D. Jennings and T. Rudolph, Phys. Rev. X {\bf 5}, 021001 (2015).
\bibitem{Kos} R. Kosloff and A. Levy, Annu. Rev. Phys. Chem. {\bf 65}, 365 (2014).
\bibitem{Gelb} D. Gelbwaser-Klimovsky, W. Niedenzu and G. Kurizki, Adv. At., Mol., Opt. Phys. {\bf 64}, 329 (2015).
\bibitem{Quan} H. T. Quan, Y. Liu, C. P. Sun and F. Nori, Phys. Rev. E {\bf 76}, 031105 (2007).
\bibitem{Roul} A. Roulet, S. Nimmrichter, J. M. Arrazola, S. Seah and V. Scarani, Phys. Rev. E {\bf 95} 062131 (2017).
\bibitem{Levy} A. Levy and R. Kosloff, Phys. Rev. Lett. {\bf 108}, 070604 (2012).
\bibitem{Mit} M. T. Mitchison, M. Huber, J. Prior, M. P. Woods and M. B. Plenio, Quantum Sci. Technol. {\bf 1}, 015001 (2016).
\bibitem{Boy} D. Boyanovsky and D. Jasnow, Phys. Rev. A {\bf 96}, 012103 (2017).
\bibitem{Erk} P. Erker P, M. T. Mitchison, R. Silva, M. P. Woods, N. Brunner and M. Huber, Phys. Rev. X {\bf 7}, 031022 (2017).
\bibitem{Hof} P. P. Hofer, M. Perarnau-Llobet, L. D. M. Miranda, G. Haack, R. Silva, J. B. Brask and N. Brunner, New J. Phys. {\bf 19}, 123037 (2017).
\bibitem{Man} S. Maniscalco, J. Piilo, F. Intravaia, F. Petruccione and A. Messina, Phys. Rev. A  {\bf 70}, 032113 (2004).
\bibitem{Gori} V. Gorini, A. Kossakowski, and E.C.G. Sudarshan, J. Math. Phys. {\bf 17}, 821 (1976).
\bibitem{Bre} H. P. Breuer and F. Petruccione, {\it The Theory of Open Quantum Systems} Oxford University Press (2002).
\bibitem{Man2} S. Maniscalco, F. Intravia, J. Piilo, and A. Messina, J. Opt. B: Quantum Semiclassical Opt. {\bf 6}, S98 (2004).
\bibitem{Schl} M. A. Schlosshauer, {\it Decoherence: and the Quantum-to-Classical Transition}, Springer Science \& Business Media (2007).
\bibitem{Sol} A. Soltanmanesh and A. Shafiee, arXiv preprint arXiv:1802.07468 (2018).
\bibitem{Lie} E. H. Lieb and M. B. Ruskai, Phys. Rev. Lett. {\bf 30}, 434 (1973).
\bibitem{Wehr} A. Wehrl, Rev. Mod. Phys. {\bf 50}, 221 (1978).
\bibitem{Abe} J. Åberg, arXiv:quant-ph/0612146 (2006).
\bibitem{Bau} T. Baumgratz, M. Cramer, and M. B. Plenio, Phys. Rev. Lett. {\bf 113}, 140401 (2014).
\bibitem{Str} A. Streltsov, G. Adesso and M.B. Plenio, Rev. Mod. Phys. {\bf 89}, 041003 (2017).
\bibitem{Win} A. Winter and D. Yang, Phys. Rev. Lett. {\bf 116}, 120404 (2016).
\bibitem{Yua} X. Yuan, H. Zhou, Z. Cao, and X. Ma, Phys. Rev. A {\bf 92}, 022124 (2015).
\bibitem{Str2} A. Streltsov, {\it Quantum Correlations Beyond Entanglement and their Role in Quantum Information Theory}, Springer Science \& Business Media (2015).
\bibitem{All3} A. Allahverdyan, R. Balian, and T. Nieuwenhuizen, Europhys. Lett. {\bf 67}, 565 (2004).
\bibitem{Pus} W. Pusz and S. L. Woronowicz, Commun. Math. Phys. {\bf 58}, 273 (1978).
\bibitem{Per} M. Perarnau-Llobet, K. V. Hovhannisyan, M. Huber, P. Skrzypczyk, J. Tura and A. Acin, Phys. Rev. E {\bf 92}, 042147 (2015).

\end{thebibliography}
\end{document}